\documentclass[epj]{svjour}

\usepackage{amsmath}
\usepackage{amssymb}
\usepackage{graphics}
\usepackage{graphicx}
\usepackage{xspace}
\usepackage{cite}
\usepackage{verbatim}
\usepackage[english]{babel}
\usepackage{hyperref}

\newcommand{\be}{\begin{equation}}
\newcommand{\ee}{\end{equation}}
\newcommand{\ben}{\begin{enumerate}}
\newcommand{\een}{\end{enumerate}}
\newcommand{\bi}{\begin{itemize}}
\newcommand{\ei}{\end{itemize}}
\newcommand{\bbe}{\begin{equation*}}
\newcommand{\eee}{\end{equation*}}
\newcommand{\bber}{\begin{equation*}\textcolor{red}}
\newcommand{\eeer}{\end{equation*}}
\newcommand{\bc}{\begin{center}}
\newcommand{\ec}{\end{center}}
\newcommand{\bea}{\begin{eqnarray}}
\newcommand{\eea}{\end{eqnarray}}
\newcommand{\bem}{\begin{pmatrix}}
\newcommand{\eem}{\end{pmatrix}}
\newcommand{\bbea}{\begin{eqnarray*}}
\newcommand{\eeea}{\end{eqnarray*}}

\newcommand{\bts}{{\emph {BtS}}\xspace}

\newcommand{\sbts}{S$_{\text{BtS}}$\xspace}
\newcommand{\snmmm}{S$_{\text{nMMM}}$\xspace}
\newcommand{\sgdas}{S$_{\text{GDAS}}$\xspace}

\newcommand{\mr}[1]{\mathrm{#1}}

\newcommand{\xmax}{{$X_\text{max}$}\xspace}

\def\EeV{\ifmmode {\mathrm{\ Ee\kern -0.1em V}}\else
                   \textrm{Ee\kern -0.1em V}\fi\xspace}%
\def\PeV{\ifmmode {\mathrm{\ Pe\kern -0.1em V}}\else
                   \textrm{Pe\kern -0.1em V}\fi\xspace}%
\def\TeV{\ifmmode {\mathrm{\ Te\kern -0.1em V}}\else
                   \textrm{Te\kern -0.1em V}\fi\xspace}%
\def\MeV{\ifmmode {\mathrm{\ Me\kern -0.1em V}}\else
                   \textrm{Me\kern -0.1em V}\fi\xspace}%
\def\GeV{\ifmmode {\mathrm{\ Ge\kern -0.1em V}}\else
                   \textrm{Ge\kern -0.1em V}\fi\xspace}%
\def\keV{\ifmmode {\mathrm{\ ke\kern -0.1em V}}\else
                   \textrm{ke\kern -0.1em V}\fi\xspace}%
\def\MeV{\ifmmode {\mathrm{\ Me\kern -0.1em V}}\else
                   \textrm{Me\kern -0.1em V}\fi\xspace}%
\def\eV{\ifmmode {\mathrm{\ e\kern -0.1em V}}\else
                   \textrm{e\kern -0.1em V}\fi\xspace}%

\def \mal      {Ma\-lar\-g\"{u}e\xspace}
\def \pao      {Pierre Auger Observatory\xspace}

\def \gcmsq    {$\text{g~cm}^{-2}$\xspace}
\def \degree   {$^{\circ}$\xspace}

\begin{document}

\title{Description of Atmospheric Conditions\\ at the \pao}
\subtitle{Using Meteorological Measurements and Models}

\author{Bianca Keilhauer\inst{1}\thanks{bianca.keilhauer@kit.edu}
        \and
        Martin Will\inst{1}\thanks{martin.will@kit.edu}
        for the Pierre Auger Collaboration\inst{2}\thanks{\url{http://www.auger.org/archive/authors_2012_06.html}}}

\institute{Karlsruher Institut f\"{u}r Technologie, Institut f\"{u}r Kernphysik, Karlsruhe, Germany
\and Observatorio Pierre Auger, Av.\ San Mart\'{\i}n Norte 304, 5613 \mal, Argentina}

\titlerunning{Description of Atmospheric Conditions Using Measurements and Models}
\authorrunning{B. Keilhauer, M. Will}

\date{Received: \today / Accepted: date}

\abstract{
  Atmospheric conditions at the site of a cosmic ray observatory must be known
  well for reconstructing observed extensive air showers, especially when
  measured using the fluorescence technique. For the Pierre Auger Observatory, a
  sophisticated network of atmospheric monitoring devices has been conceived.
  Part of this monitoring was a weather balloon program to measure atmospheric
  state variables above the Observatory. To use the data in reconstructions of
  air showers, monthly models have been constructed. Scheduled balloon launches
  were abandoned and replaced with launches triggered by high-energetic air
  showers as part of a rapid monitoring system. Currently, the balloon launch
  program is halted and atmospheric data from numerical weather prediction
  models are used. A description of the balloon measurements, the monthly models
  as well as the data from the numerical weather prediction are presented.
}

\maketitle

\section{Introduction
\label{sec:intro}}

At the \pao, extensive air showers induced by cosmic rays with energies above
$10^{17}$~eV are measured. With increasing energy of the primary particle, the
flux of cosmic rays is strongly decreasing and it is expected that a transition
from cosmic rays with galactic origin to those with extragalactic origin is
occurring in the observed energy range of the \pao. Besides the energy of the
cosmic rays, which is needed to evaluate the spectral slope of the cosmic ray
flux, another information of utmost importance is their mass composition. For
the detection of extensive air showers with high statistics, a surface detector
array consisting of more than 1\,600 water Cherenkov tanks is installed at the
Pampa Amarilla close to \mal, Argentina. The second component of the Observatory
are 27 fluorescence telescopes at four sites observing faint nitrogen
fluorescence in the atmosphere above the ground array. The fluorescence emission
is induced by the giant cascade of secondary particles developing after the
primary cosmic ray interacts with nuclei in the upper parts of the atmosphere.
This dim light can only be detected during nights with low atmospheric opacity
and little light background from the moon or other sources.

The fluorescence technique is very susceptible to changes in atmospheric state
variables and optical properties of the air. Both depend on air temperature,
pressure and on water vapor pressure. For this reason, the \pao employs a
sophisticated atmospheric monitoring program~\cite{Abraham:2010} to measure
atmospheric properties. This includes infrared cloud cameras and lidar stations
at every FD site~\cite{Tonachini:2012lid} to scan for clouds, a weather balloon
program~\cite{BenZvi:2012xts} to measure profiles of atmospheric state variables
and five weather stations distributed over the array of the Auger Observatory to
monitor surface values. The lidar stations and two central laser facilities are
used to measure the aerosol optical depth~\cite{Wiencke:2012epj}. All this
information is adequately processed to be included in the reconstruction
procedure for measured extensive air showers.  Critical aspects are the spatial
and temporal resolution of the locally observed atmospheric conditions. In case
of profiles of atmospheric state variables, monthly models were compiled using
radiosonde data~\cite{BenZvi:2012xts}. More recently, the applicability of data
from the Global Data Assimilation System (GDAS) for the \pao has been
investigated~\cite{Abreu:2012gdas}.

The paper is structured as follows: In Sec.~\ref{sec:fluo}, we will give a short
introduction to expected effects of atmospheric properties on the development
and detection of extensive air showers. The locally observed data of state
variables by weather balloons and their application in the air shower
reconstruction are described in Sec.~\ref{sec:balloon}. In Sec.~\ref{sec:nwp},
GDAS data are described and compared to local measurements at the Auger
Observatory and the development of monthly models from these data is detailed. A
brief discussion of uncertainties within the air shower reconstruction caused by
variations of atmospheric conditions is given in Sec.~\ref{sec:reco}.

\section{Impact of Atmospheric State Variables on Air Shower Measurements
\label{sec:fluo}}

Two important aspects of air shower reconstruction are the determination of the
energy of the primary particle of each shower and the estimate of the mass
composition for a large set of observed events. The first quantity is obtained
at the \pao by an almost calorimetric measurement of the energy deposited in the
atmosphere during the air shower development~\cite{Abraham:2009pm}. The induced
fluorescence light is proportional to the energy deposit at each state of shower
development~\cite{Ave:2007b}. The integral detection of the light trace by
fluorescence telescopes enables a very precise determination of the energy of
the cosmic ray particle. The second quantity, the mass composition of air
showers, is derived by investigating the position of the shower maximum \xmax,
the altitude given in units of \gcmsq where the number of secondary particles is
highest~\cite{Abraham:2010b}. For the general case of vertical incidence and
equal energy of the primary cosmic ray particle, nuclei of light elements
initiate deeply penetrating air showers and more shallow showers are formed by
interactions of heavier elements. \xmax is fluctuating considerably from shower
to shower, so large statistics is needed to draw conclusions about mass
composition. Larger fluctuations of \xmax are observed for lighter elements than
for elements with a high number of nuclei. With fluorescence telescopes, the
atmospheric depth of shower maximum can be measured directly.

Atmospheric state variables, mainly the density of air $\rho$, alter the
development of extensive air showers in the atmosphere as it is governed by the
interactions and decays of the secondary particles. These processes are
determined by the atmospheric depth $X$, which is the path integral of $\rho$
from the top of the atmosphere to a given altitude. Since the fluorescence
telescopes observe the light trace with respect to geometrical altitude $h$, the
conversion between $h$ and $X$ depends on atmospheric conditions.

Most sensitive to short-term variations of state variables is the fluorescence
yield of nitrogen in
air~\cite{Arqueros:2008,Keilhauer:2008,Arqueros:2011,Keilhauer:2012}. The main
emission bands for nitrogen molecules are in the range between about 280 and
420\,nm. The energy deposit of the air shower in the atmosphere causes an
excitation of mainly the second positive system of N$_2$ and the first negative
system of N$_2^+$. The spontaneous de-excitation, yielding the fluorescence
light emission, is counteracted by collisional de-excitation of nitrogen
molecules by other nitrogen or oxygen molecules and by water vapor. These
quenching processes are dependent on atmospheric conditions. The rate of
collisions between the molecules can be described by kinetic gas theory,
including temperature and pressure dependence. In addition, the collisional
cross sections between nitrogen and air molecules in general were measured to be
explicitly temperature-dependent~\cite{Ave:2008,Pereira:2010}. The content of
water vapor in air adds another quenching source.

All these aspects require a sophisticated consideration of actual atmospheric
conditions in the process of air shower reconstruction. Thus, the \pao is
operating a program for measuring profiles of atmospheric state variables with
meteorological radio soundings supplemented by a set of ground-based weather
stations.

\section{The Weather Balloon Program
\label{sec:balloon}}

In the process of air shower reconstruction, the observed fluorescence light at
the telescopes is interpreted as energy deposited by the air shower in the
atmosphere along the shower trajectory. This requires intermediate steps of
correcting the observed light by atmospheric attenuation (absorption and
scattering out of the field of view) and of determining the effective
fluorescence yield taking into account all quenching processes. The atmospheric
properties needed for fluorescence description are described in
Sec.~\ref{sec:fluo}. In the case of ground-based fluorescence telescopes, like
the detectors used at the Auger Observatory, light absorptions by atmospheric
trace gases is of minor importance. More relevant are aerosol and molecular
scattering. Aerosol conditions above the Observatory are measured every hour
during data taking of the telescopes~\cite{Wiencke:2012epj}. The molecular
scattering, i.\,e. Rayleigh scattering, depending on temperature, pressure and
humidity, is calculated during reconstruction by applying atmospheric profiles
of these quantities.

\subsection{Radiosoundings
\label{sec:radio}}

The most direct way to obtain altitude-dependent profiles of state variables is
given by measurements with meteorological radiosondes launched on weather
balloons. At the \pao, intermittent launches were performed between August 2002
and December 2010. During the first years, campaigns of about 3 weeks with an
average of 9 launches per campaign were done roughly 3 times per year. Starting
in July 2005, the rate of launches became more regularly to about one launch
every five days. In March 2009, the radio soundings were implemented into the
rapid atmospheric monitoring program of the Auger
Observatory~\cite{BenZvi:2012xts}, triggering a launch of a weather balloon
shortly after the detection of particularly interesting air showers like very
high-energetic events, see Sec.~\ref{sec:bts}. In total, 331 successfully
measured profiles are gathered until end of 2010.

\begin{figure*}[htbp]
  \begin{center}
  \begin{minipage}[t]{.49\textwidth}
    \centering
    \includegraphics*[width=.99\linewidth,clip]{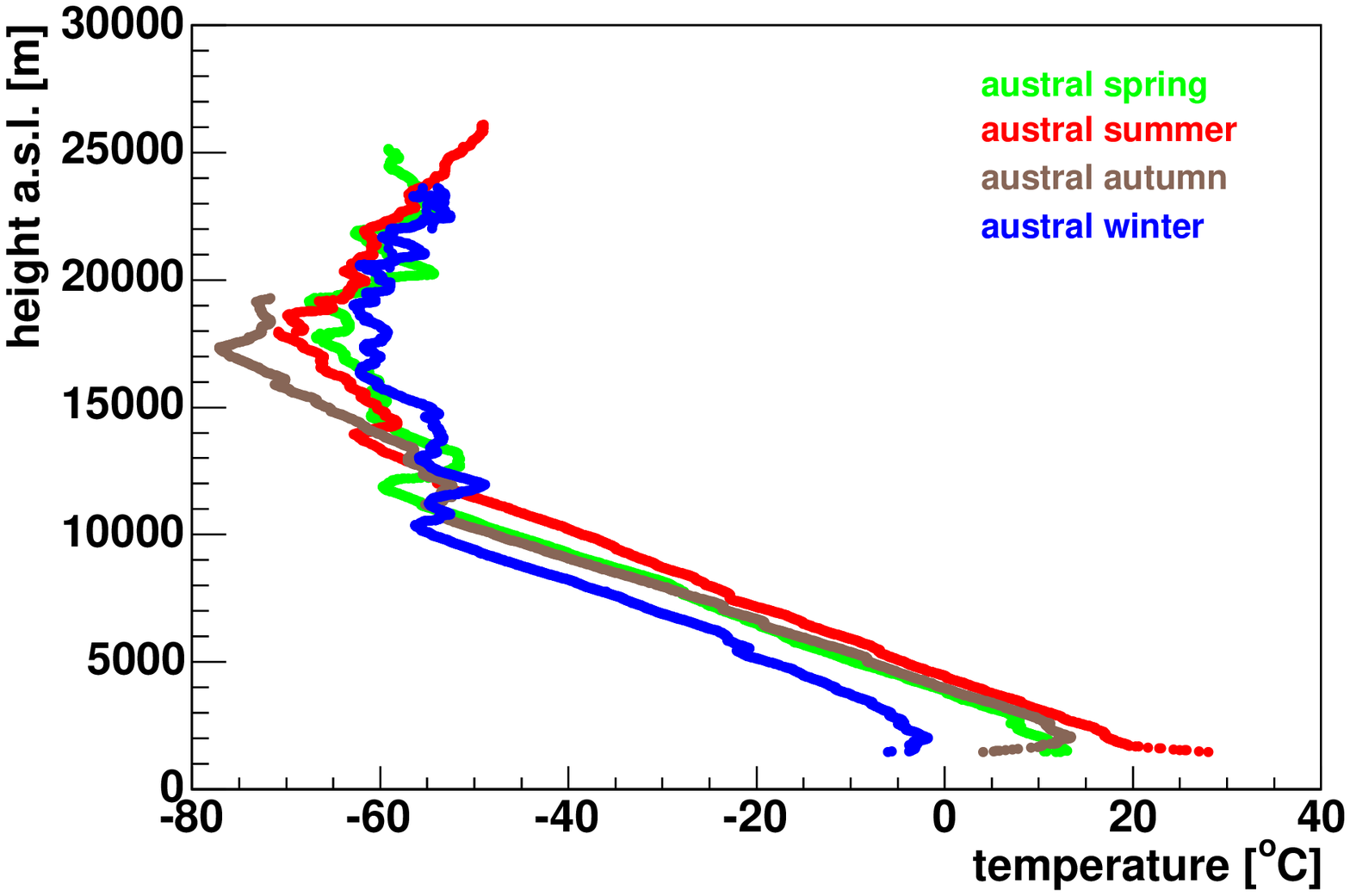}
    \includegraphics*[width=.99\linewidth,clip]{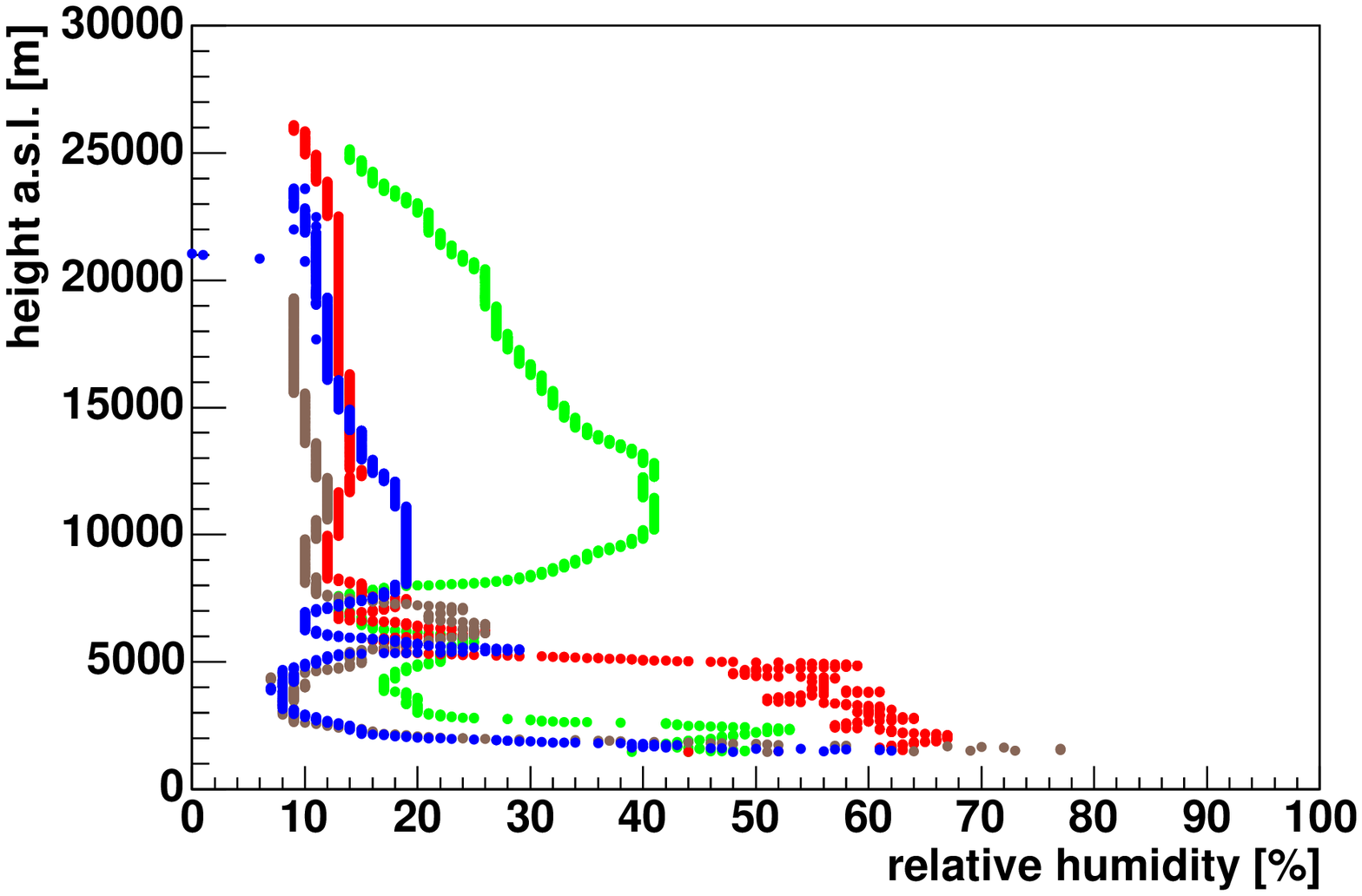}
  \end{minipage}
  \hfill
  \begin{minipage}[t]{.49\textwidth}
    \centering
    \includegraphics*[width=.99\linewidth,clip]{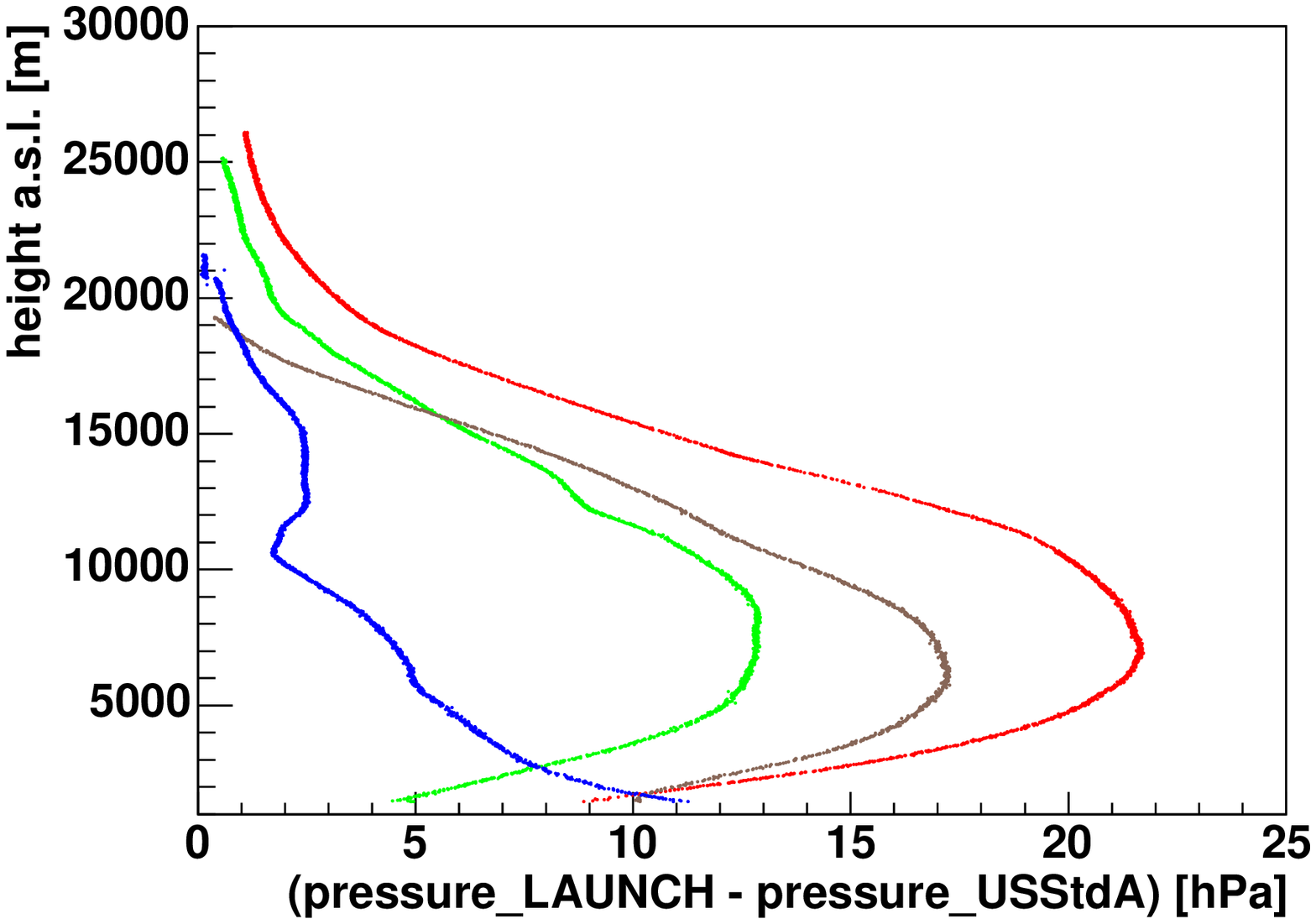}
    \includegraphics*[width=.99\linewidth,clip]{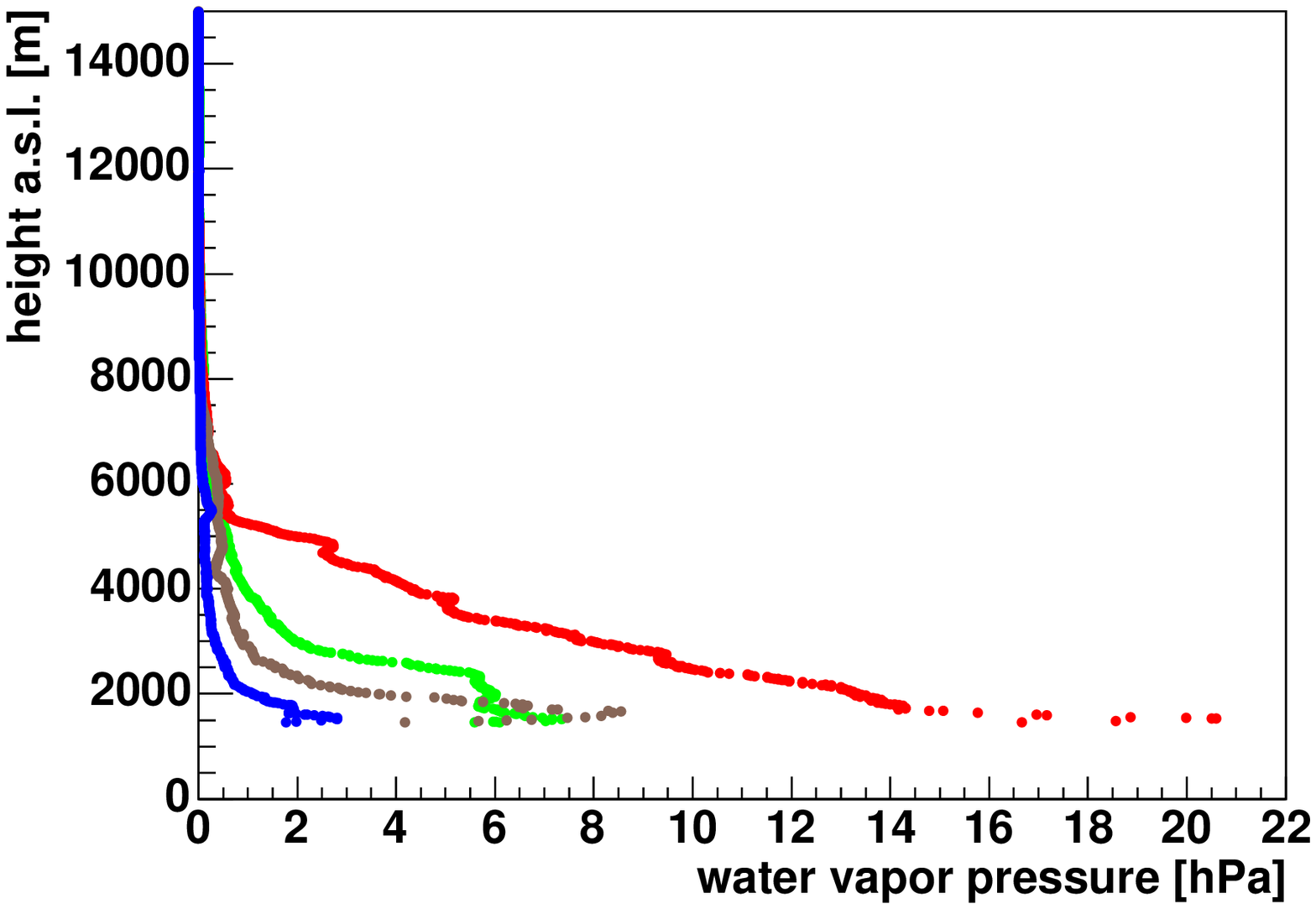}
  \end{minipage}
  \begin{minipage}[c]{.6\textwidth}
    \centering
    \includegraphics*[width=.99\linewidth,clip]{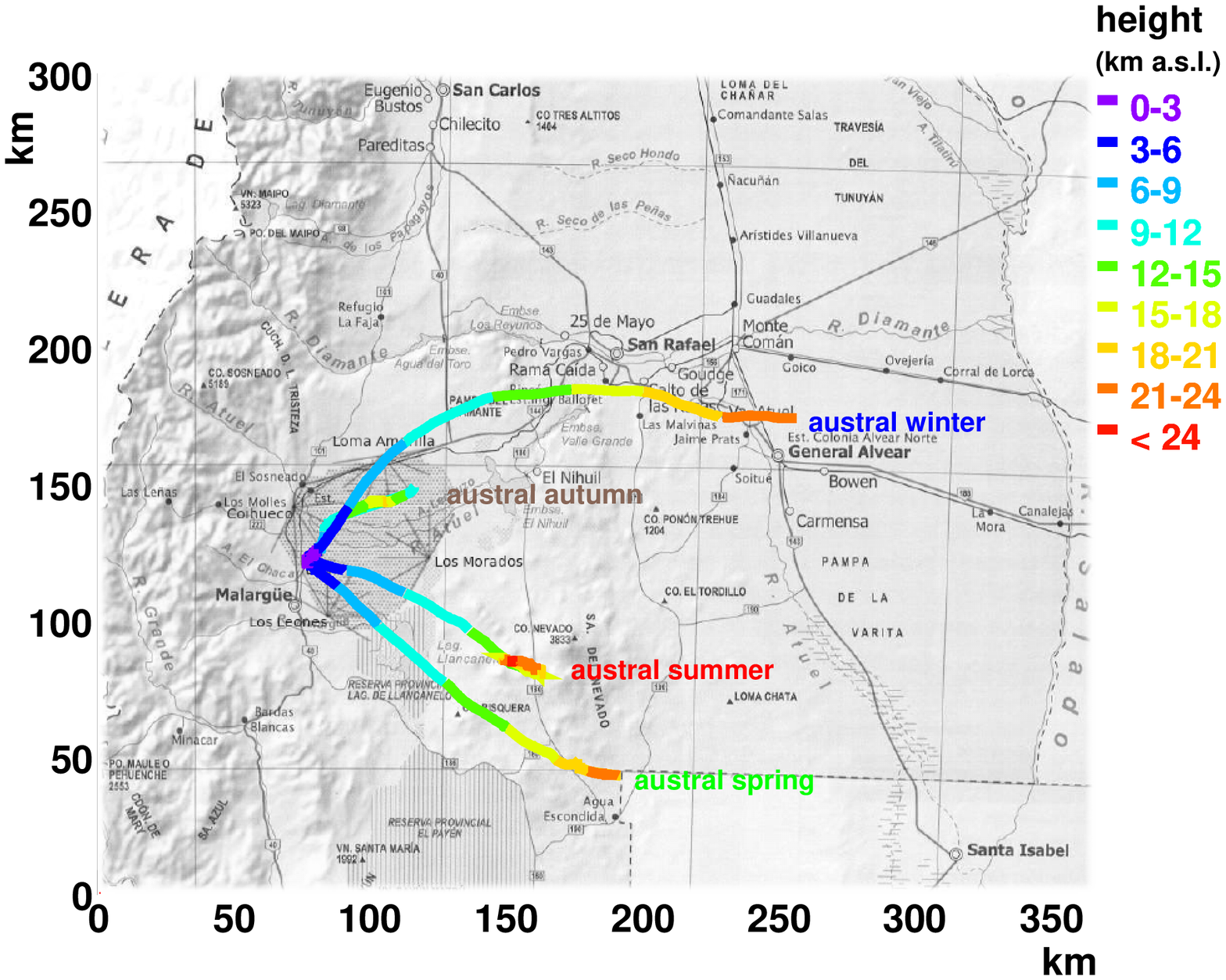}
  \end{minipage}
  \caption[]{\label{fig:launches}
    One characteristic atmosphere per season obtained by meteorological
    radiosondes above the array of the \pao. The profiles for spring were
    measured at Oct., 3rd, 2007, 9:17 UTC (start time of weather balloon), for
    summer at Feb., 19th, 2008, 00:29 UTC, for autumn at Apr., 28th, 2007, 9:26
    UTC, and for winter at Aug., 10th, 2007, 9:26 UTC. Top left: temperature,
    Top right: difference of measured pressure profile to that of the US
    Standard Atmosphere 1976 (USStdA)~\cite{USStdA:1976}, Bottom left: relative
    humidity, Bottom right: water vapor pressure.  Bottom: Flight path of
    weather balloons\setcounter{footnote}{1}\footnotemark[\value{footnote}].
  }
  \end{center}
\end{figure*}

With the used radiosondes~\cite{graw}, air temperature $T$ and relative humidity
$u$ are measured directly via dedicated sensors. Until October 2008, a type of
radiosonde was used with active measurement of the pressure $p$. With the more
recent type of sonde, the pressure is calculated iteratively from ground
pressure, the measured temperature profile and altitude from GPS information.
The wind speed and direction is obtained from the GPS position as well. All
derived variables like water vapor pressure and air density are calculated
afterwards. In Fig.~\ref{fig:launches}, one example measurement of the
atmospheric conditions per season is displayed. Because of the exponential
decrease of $p$, this variable is plotted as difference to the US Standard
Atmosphere 1976 (USStdA)~\cite{USStdA:1976}. Significant differences to the
USStdA are found between about 5 and 12~km~a.s.l.\ where the most dynamic
development of extensive air showers is taking place. Air humidity is relevant
only in the lowest 6~km~a.s.l., influencing mainly the air fluorescence yield
(see Sec.~\ref{sec:fluo}). The main wind direction is west at the site of the
Auger Observatory, only few balloons were driven by east wind.  As indicated by
the altitude color code\setcounter{footnote}{0}\footnote{For interpretation of
the colors in this figure, the reader is referred to the web version of this
article.} in the bottom panel of Fig.~\ref{fig:launches}, most of the weather
balloons were still above the array of the Observatory until about 10~km~a.s.l.,
the conditions were indeed measured at the place of interest.

\subsection{Monthly Models
\label{sec:nmmm}}

Radiosonde launches are the most accurate way to determine the current profiles
of atmospheric state variables for air shower reconstruction. Meteorological
institutions performing synoptic measurements around the world launch at least
two balloons per day, while at the \pao, only one balloon every few days was
launched. Even though the atmospheric conditions above the Pampa Amarilla vary
only moderately, it is not possible to simply interpolate between two launches
for the application in air shower reconstruction. To cover all periods and in
particular the intermediate time periods, monthly models have been constructed
for the site of the Observatory. These models have been compiled two times from
locally measured weather data, once in 2005 and an updated set was produced in
2009. The models from 2005 are mostly based on radiosonde measurements from
C\'{o}rdoba and Santa Rosa, Argentina, both more than 500\,km away from \mal.
They are adjusted in the lower part of the atmosphere by local radiosonde
launches performed above the \pao. 

Until December 2008, 277~launches were performed on site, so models based solely
on local measurements could be produced. Unlike the previous version, this new
set also includes air humidity. All data measured during day and night are used,
but some criteria are applied to the dataset to avoid biasing the monthly
models. If there are more than one launch performed within one day or within one
night, only one data set is selected for the models. Therefore, only one launch
per nine hours is considered. This leaves data from 261~launches. For the
monthly vapor pressure models, 32 more launches during overcast conditions are
excluded.

\begin{figure*}[tbp]
  \begin{center}
  \begin{minipage}[t]{.49\textwidth}
    \centering
    \includegraphics*[width=\linewidth,height=0.24\textheight]{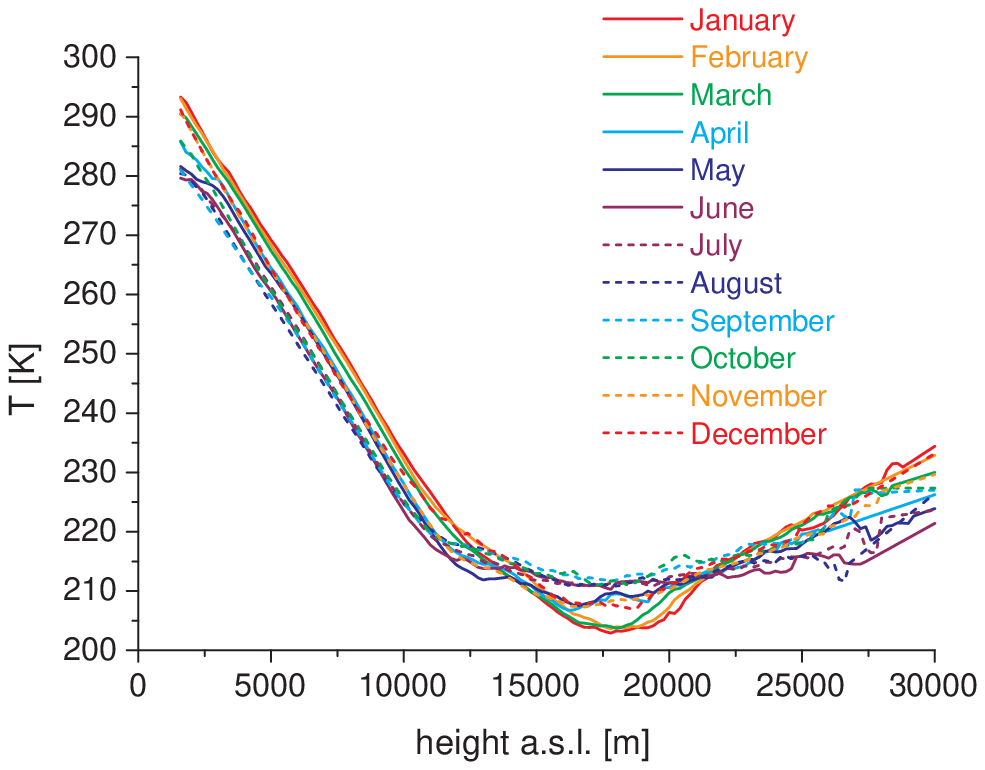}
  \end{minipage}
  \hfill
  \begin{minipage}[t]{.49\textwidth}
    \centering
    \includegraphics*[width=\linewidth,,height=0.24\textheight]{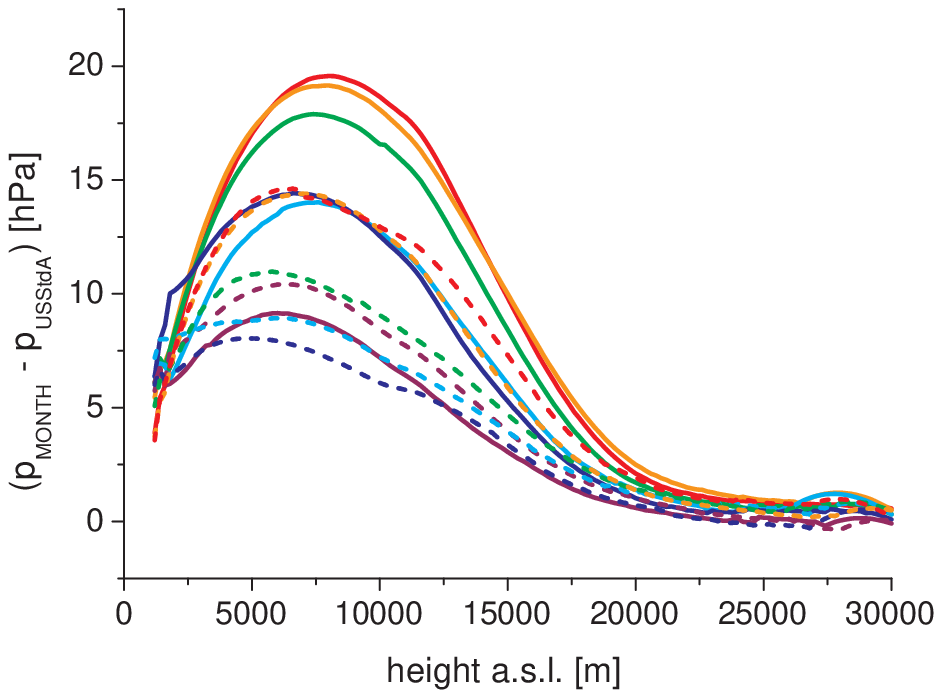}
  \end{minipage}
  \begin{minipage}[t]{.49\textwidth}
    \centering
    \includegraphics*[width=\linewidth,,height=0.24\textheight]{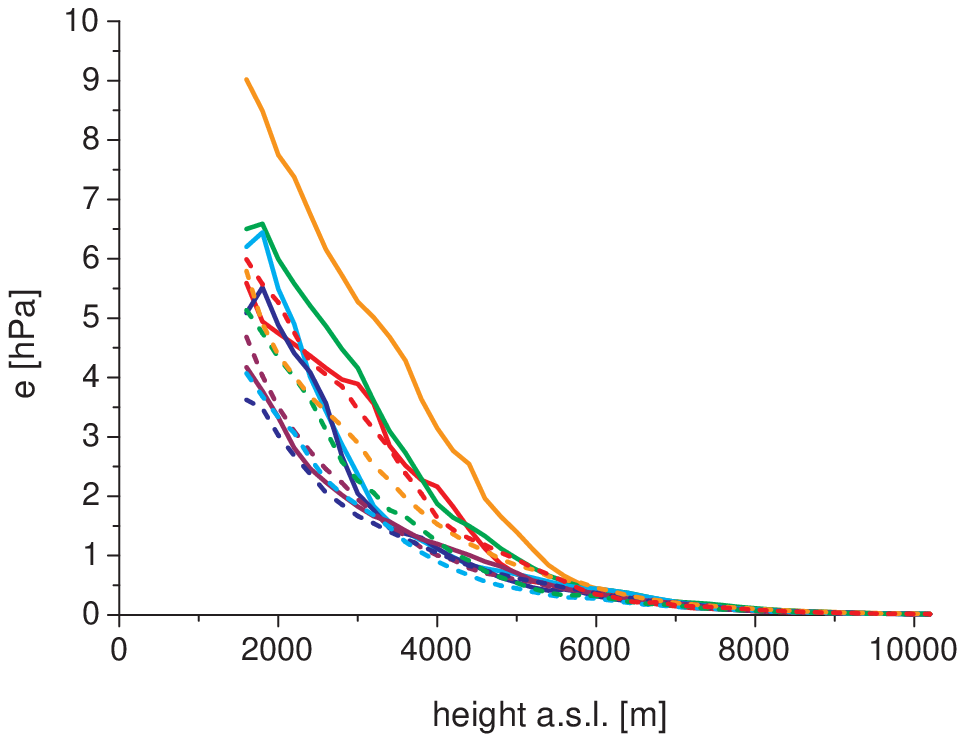}
  \end{minipage}
  \hfill
  \begin{minipage}[t]{.49\textwidth}
    \centering
    \includegraphics*[width=\linewidth,,height=0.24\textheight]{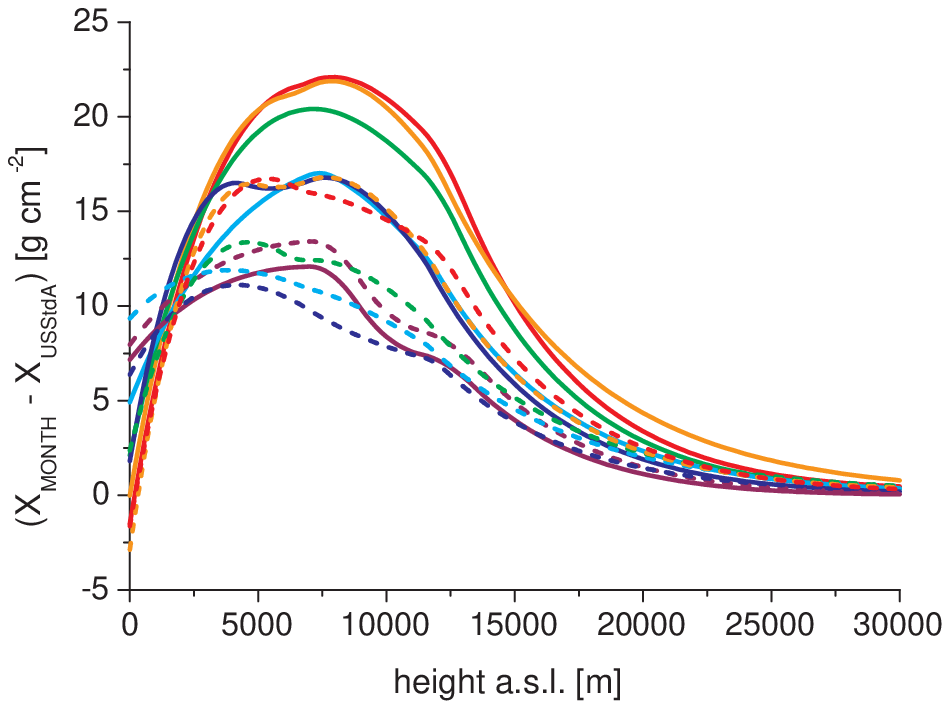}
  \end{minipage}
  \caption{\label{fig:nMMM}
    The \mal Monthly Models for temperature, pressure, water vapor pressure and
    atmospheric depth derived from local measurements until end of 2008. 
    For pressure and atmospheric depth, the difference to the USStdA is shown.
  }
  \end{center}
\end{figure*}

For each launch, measured temperature, pressure and relative humidity profiles
are available. Vapor pressure and density are then calculated using the Magnus
formula and the ideal gas law. Profiles of the atmospheric depth are calculated
by integrating the density profile from the height at balloon burst $h_\mr{b}$
down to the ground. The initial atmospheric depth at burst height is estimated
by the pressure and the gravitational acceleration, $X(h_\mr{b}) = p(h_\mr{b}) /
g(\Phi,h_\mr{b})$, with $\Phi$ being the geographical latitude. Typical burst
heights are between 20 and 25\,km corresponding to atmospheric depths between 20
to 40\,\gcmsq. All profiles of each month are then averaged giving profiles
starting at 1\,600~m~a.s.l. Using data from five weather stations across the
Observatory, the averages are extrapolated down to 1\,200\,m~a.s.l. The monthly
profiles for temperature $T$ and water vapor pressure $e$ are displayed in
Fig.~\ref{fig:nMMM}, along with the differences of pressure $p$ and atmospheric
depth $X$ to the profiles of the USStdA.

For an appropriate use in air shower simulations, e.\,g.\ using
CORSIKA~\cite{Heck:1998}, every monthly atmospheric depth average is fitted with
an exponential function for four different layers $i$,
\begin{equation}
\label{eq:Xparam}
  X(h) = a_i + b_i e^{-h/c_i}.
\end{equation}
The four sets of parameters $a_i$, $b_i$, $c_i$ and the heights of the
boundaries between the layers are fit taking into account given boundary and
continuity conditions between the layers. The local models reach up to 30~km,
but for air shower simulations, data are required until the top of the
atmosphere. This is achieved by applying adjusted density data from the USStdA
above the local profiles up to 100~km. Above, the atmospheric depth is described
by a linear decrease to Zero which is reached at 112.8~km~a.s.l. This fifth
layer is adopted from the standard parameterization of the USStdA which is
available in CORSIKA.

\subsection{The Balloon-the-Shower program
\label{sec:bts}}

The rapid atmospheric monitoring program is conceived to provide almost
real-time atmospheric data for air shower events that are of special interest to
the analyses of the \pao~\cite{BenZvi:2012xts}. This includes events with very
high energies that are used for the energy calibration of the SD
array~\cite{Pesce:2011icrc} as well as anisotropy and composition
studies~\cite{Abraham:2007,Abraham:2010yv}. Also, showers with distorted
profiles are of interest, since they might arise from exotic particles or
unusual hadronic interactions~\cite{Baus:2011icrc}. To be able to trigger a
dedicated atmospheric measurement, air shower data are put to an online
reconstruction within a few minutes after they are gathered in the central
campus of the Observatory.

The balloon program became part of this program in March 2009 to reduce the time
after an high-energetic air shower from up to several days down to less than
three hours. Relevant air shower parameters, like reconstructed energy, shower
profile, detector information and uncertainty estimates, from the online
reconstruction are sent to a computer running an analysis script that checks for
new showers every 15~minutes for profiles that meet certain quality criteria. In
order to reduce the number of possible air showers triggering the \bts
(Balloon-the-Shower) program, an energy threshold of 10$^{19.3}$\,eV (about
20\,EeV) was implemented.

Once an interesting shower is found, a local technician is informed via a short
text message. He then drives to the Balloon Launching Station (BLS,
c.\,f.~Fig.~\ref{fig:gridPoints}) close to the south-western boundary of the
surface detector array to launch a probe. Radiosondes are launched only within 3
hours after the time of an air shower event. During the run of \bts between
March 2009 and December 2010, 100 interesting showers were identified, followed
by sending a text message, and yielding 52 successfully launched balloons.
During some launches, further triggers were registered, so that a total number
of 62 air shower events are covered by actual radio soundings. The remaining
triggers were lost due to diverse technical failures.

An analysis of reconstructed shower parameters with profiles measured within the
\bts program instead of monthly models shows a clear improvement of the accuracy
of the reconstruction and no systematic effect due to shower geometry or time of
year are found, see Sec.~\ref{sec:reco}. The \bts data played an important part
in validating new atmospheric model data for the air shower reconstruction at
the \pao. The on-site measurements served as a reference for investigating the
applicability of data from the Global Data Assimilation System (GDAS) for the
Observatory, see next Section.

\section{Data From Numerical Weather Prediction
\label{sec:nwp}}

Nowadays, numerical weather prediction is based on the process of \emph{data
assimilation}. Using complex mathematical models, the atmospheric conditions can
be predicted at all positions worldwide. To consolidate these models, the
predictions are supplemented with measured data from weather stations, weather
balloon launches, measurements from aircraft and ships, as well as satellite
data.

\subsection{The Global Data Assimilation System
\label{sec:gdas}}

\begin{figure*}[t]
  \begin{center}
  \begin{minipage}[c]{.64\textwidth}
    \centering
    \includegraphics[width=0.95\linewidth]{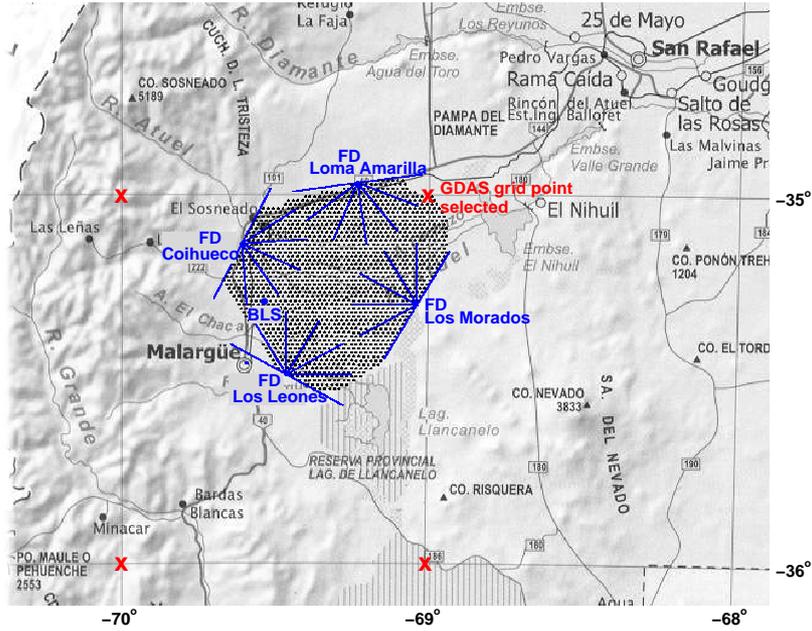}
    \caption{\label{fig:gridPoints}
      Map of the region around the Pierre Auger Observatory with geographical
      latitudes and longitudes marked at the boundary of the figure. The
      positions of the surface and fluorescence detectors are superimposed. The
      GDAS grid points closest to the Observatory are marked as red crosses.
      The coordinates of \mal are $-35.48^{\circ}$ (south) and $-69.58^{\circ}$
      (west).
    }
  \end{minipage}
  \end{center}
\end{figure*}

The Global Data Assimilation System (GDAS) is provided by the American National
Oceanic and Atmospheric Administration (NOAA). The data are available in a
resolution of three hours and a 1\degree latitude-longitude grid for the entire
globe without charge via NOAA's Real-time Environmental Applications and Display
sYstem (READY)~\cite{GDASinformation}. In Fig.~\ref{fig:gridPoints}, the
available grid points close to the Auger array are shown. Superimposed are the
positions of the regular SD grid and the location and field of views of the four
FD sites. The two grid points to the West of the Observatory are already in the
Andes mountains and therefore not suited. The point to the North-East was chosen
as it is the closest to the array. The differences between this point and the
grid point to the South-East are on average less than 1~$^\circ$C in temperature
and less than 0.3~hPa in water vapor pressure at all altitudes, making an
extrapolation unnecessary.

GDAS data is provided on 23~pressure surfaces ranging from 1000\,hPa (sea level)
to 20\,hPa (about 26\,km a.s.l.). Additionally, surface values are included. For
the site of the Auger Observatory, the lowest 5~pressure levels are below the
surface height. This is a remnant of the mathematical nature of the numerical
weather prediction model and the data of these levels are discarded. Useful data
for the Auger site are available starting in June 2005.

\begin{figure*}[htbp]
  \begin{center}
  \begin{minipage}[t]{.49\textwidth}
    \centering
    \includegraphics*[width=.9\linewidth,clip]{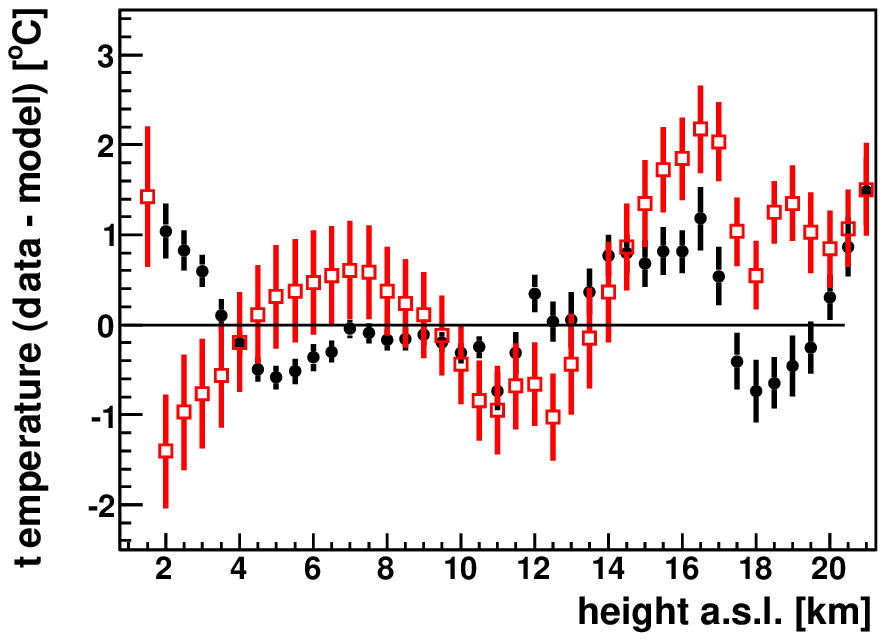}
  \end{minipage}
  \hfill
  \begin{minipage}[t]{.49\textwidth}
    \centering
    \includegraphics*[width=.9\linewidth,clip]{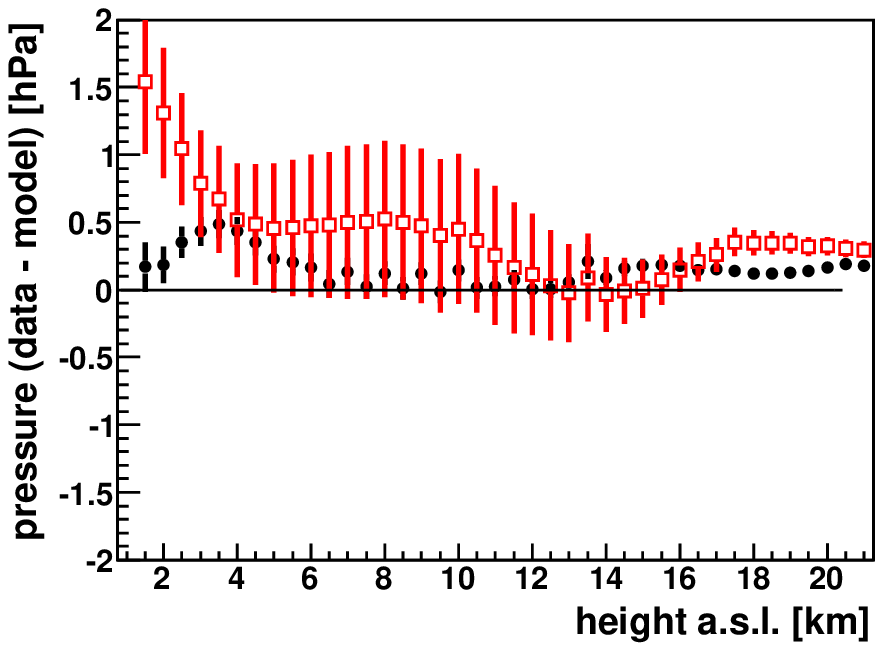}
  \end{minipage}
  \begin{minipage}[t]{.49\textwidth}
    \centering
    \includegraphics*[width=.9\linewidth,clip]{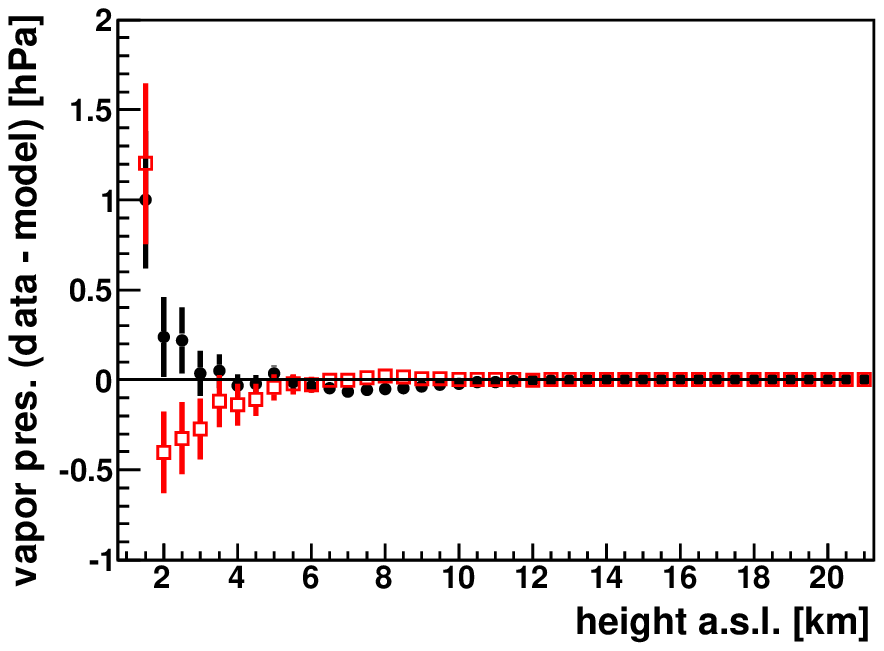}
  \end{minipage}
  \hfill
  \begin{minipage}[t]{.49\textwidth}
    \centering
    \includegraphics*[width=.9\linewidth,clip]{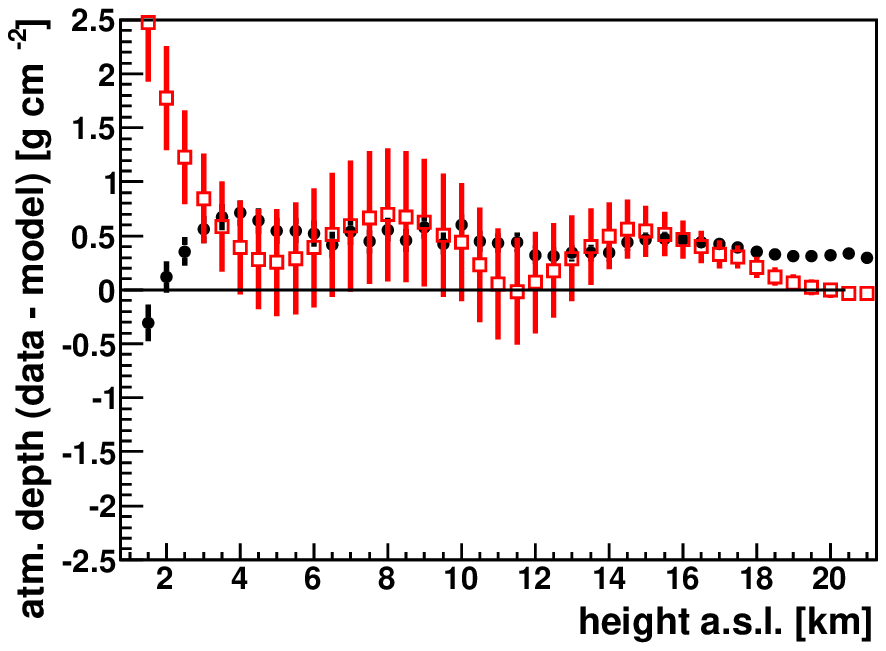}
  \end{minipage}
  \caption{\label{fig:SondeVsGDAS_nMMM_2009}
    Average difference of measured radiosonde data and GDAS models (black dots)
    and monthly mean profiles (red squares) versus height for all radiosondes
    performed in 2009 and 2010. Error bars denote the RMS spread of the
    difference.
  }
  \end{center}
\vspace{12pt}
  \begin{center}
  \begin{minipage}[t]{.32\textwidth}
    \centering
    \includegraphics*[width=.99\linewidth,clip]{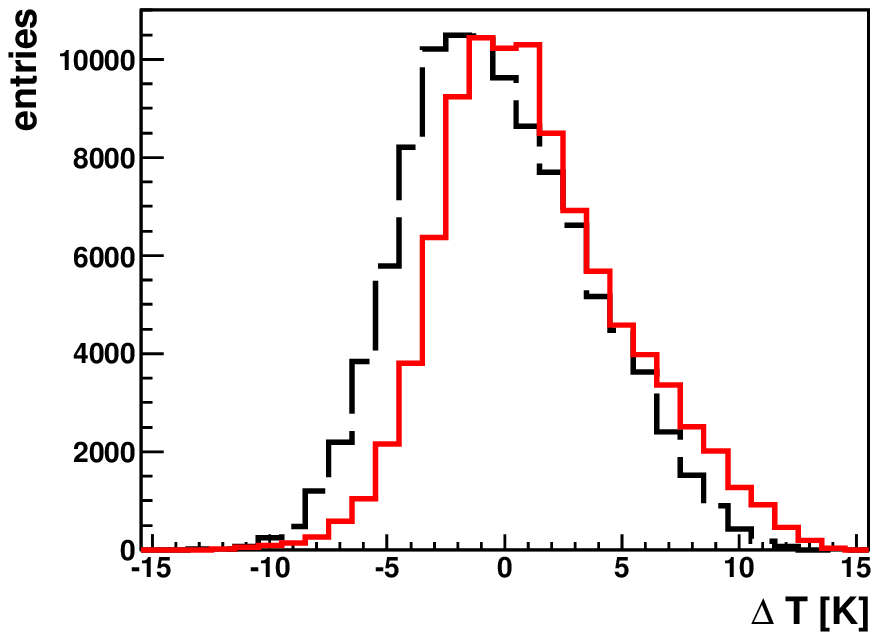}
  \end{minipage}
  \begin{minipage}[t]{.32\textwidth}
    \centering
    \includegraphics*[width=.99\linewidth,clip]{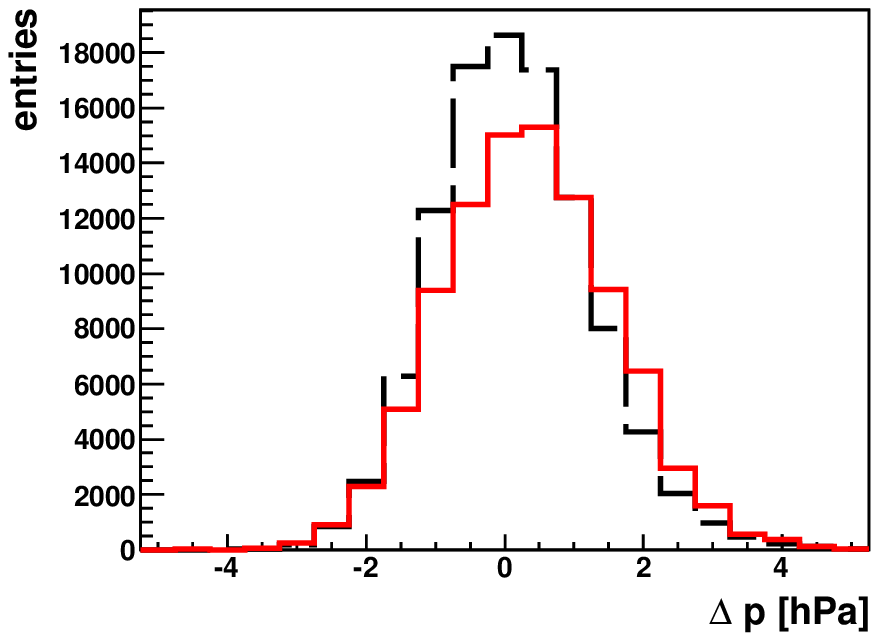}
  \end{minipage}
  \begin{minipage}[t]{.32\textwidth}
    \centering
    \includegraphics*[width=.99\linewidth,clip]{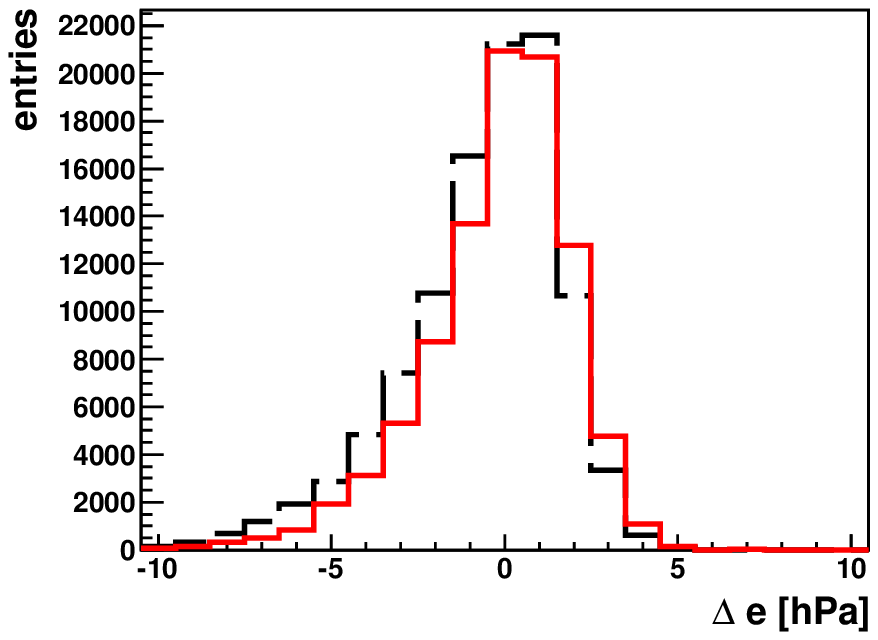}
  \end{minipage}
  \hfill
  \caption{\label{fig:WSvsGDAS}
    Difference of measured weather station data and GDAS models for all data
    from 2009. The differences GDAS model minus measured data in temperature
    $T$, pressure $p$, and water vapor pressure $e$ are shown for the weather
    station at the CLF (solid red line) and the station at FD Loma Amarilla
    (dashed black line).
  }
  \end{center}
\end{figure*}

To check the applicability of the GDAS data at the Auger site, they were
compared with the local radiosonde data and weather station data on the ground.
In Fig.~\ref{fig:SondeVsGDAS_nMMM_2009}, for several altitudes the differences
in temperature, pressure, water vapor pressure and atmospheric depth between
measured radiosonde data and the corresponding data from the GDAS model (black
dots) and the \mal monthly models (red open squares) are shown for measurements
in 2009 and 2010. This time period was chosen since the monthly models were
compiled from data until the end of 2008, so the balloon data from 2009 and 2010
represent an independent data sample to the monthly models. The error bars
denote the RMS of the differences at each height. Closer to the ground, slightly
larger differences are found in all quantities. In all four quantities, both
models describe the local conditions well. However, it is clear that the GDAS
model overall shows slightly smaller differences than the monthly models and,
more importantly, has much smaller deviations between individual profiles as
indicated by the much smaller error bars. It should be mentioned that the
agreement between local radio soundings and the monthly models is similarly well
as the GDAS model comparison with local measurements for the years 2005 to
2008~\cite{Abreu:2012gdas}, but still GDAS has a smaller spread.

The surface values of the GDAS data are investigated by comparisons with data
from the weather stations. In Fig.~\ref{fig:WSvsGDAS}, the differences of
measured weather station data and GDAS data are shown for two weather stations
(CLF and Loma Amarilla) at the time of the GDAS data set. The GDAS data are
interpolated at the height of the respective weather station and agree well with
the weather station data. The mean deviation of the distributions is comparable
to the difference seen between individual weather stations across the array.

In summary, the GDAS data describe the conditions at the \pao very well as shown
by comparisons with locally measured data. With its superior temporal
resolution, they are an excellent source for profiles of atmospheric state
variables. Furthermore, the GDAS data sets contain a lot more information that
might be explored in the future.

\subsection{Monthly Models from GDAS Data
\label{sec:gdasmon}}

The 3-hourly GDAS data are very well suited for data reconstruction at the \pao,
as shown in the previous section. For air shower simulations, e.\,g.\ with
CORSIKA, a parameterized atmospheric depth is needed. This is a complex
procedure and cannot be repeated every three hours for GDAS data, resulting in
an excessive number of possible atmospheric conditions to choose from during
simulations. Therefore, GDAS data are not applicable for simulation. Monthly
models are a good compromise between mapping seasonal atmospheric variations on
air shower development and exploiting the good agreement of GDAS data at the
Auger site. Averaging the GDAS data in monthly bins for the time period of June
2005 until May 2011 shows clear seasonal trends at all altitudes for
temperature, pressure, water vapor pressure, air density and atmospheric depth.

After creating monthly averages of all relevant atmospheric state variables, the
atmospheric depth is fit with an exponential function, see
Eq.~\eqref{eq:Xparam}, and for all 12 months, the atmospheric depths profiles
are parameterized as described in Sec.~\ref{sec:nmmm}. The parameterized
atmospheric depth functions describe the averaged depth profiles very well, only
showing small differences of less than 0.5\,\gcmsq at all altitudes. These
residuals are much smaller than those for the monthly models because of the
rather small number of local measurements used to develop the models as
described in Sec.~\ref{sec:nmmm}.

\begin{figure*}[t]
  \begin{center}
  \begin{minipage}[t]{.49\textwidth}
    \centering
    \includegraphics*[width=.9\linewidth,clip]{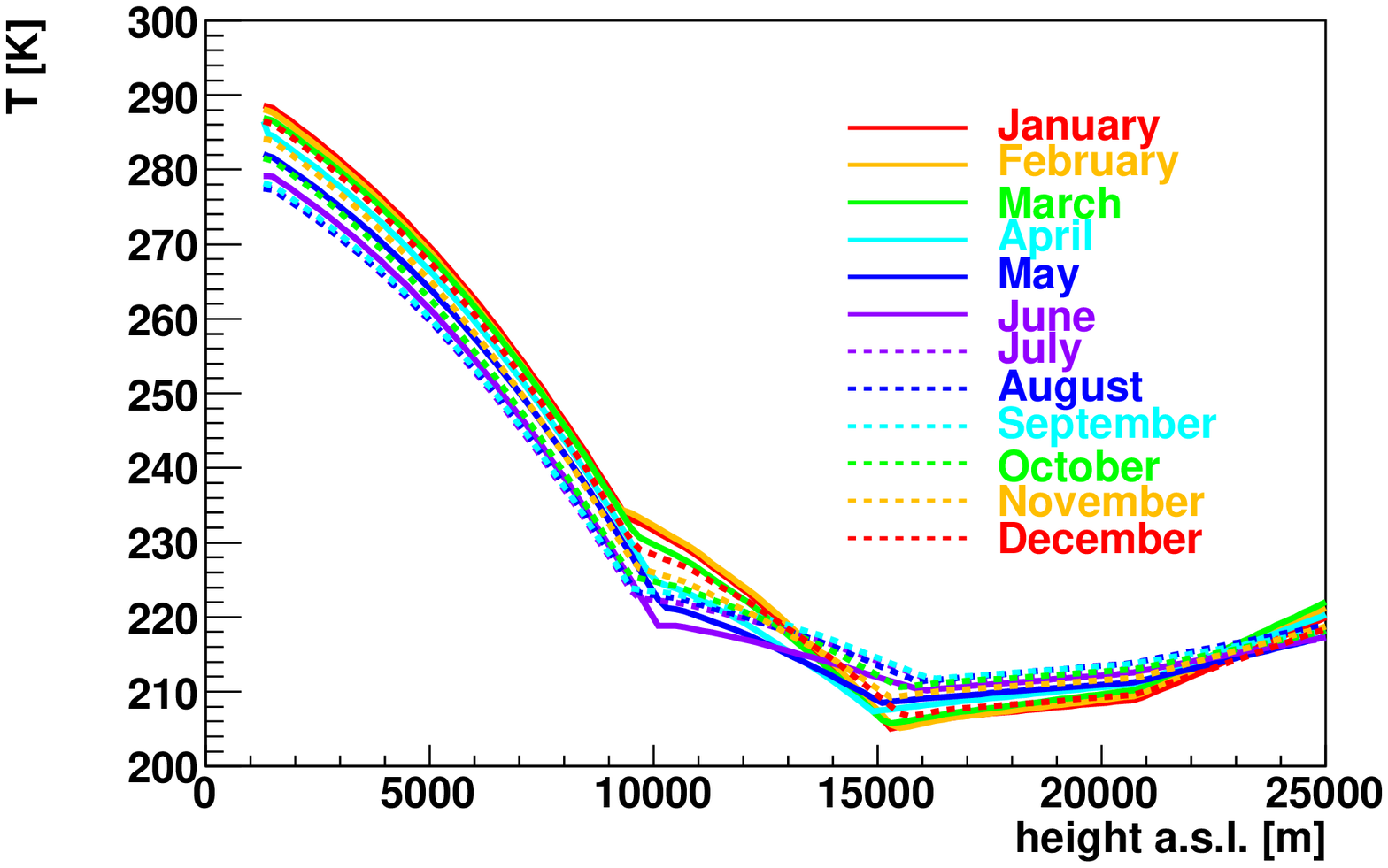}
  \end{minipage}
  \hfill
  \begin{minipage}[t]{.49\textwidth}
    \centering
    \includegraphics*[width=.9\linewidth,clip]{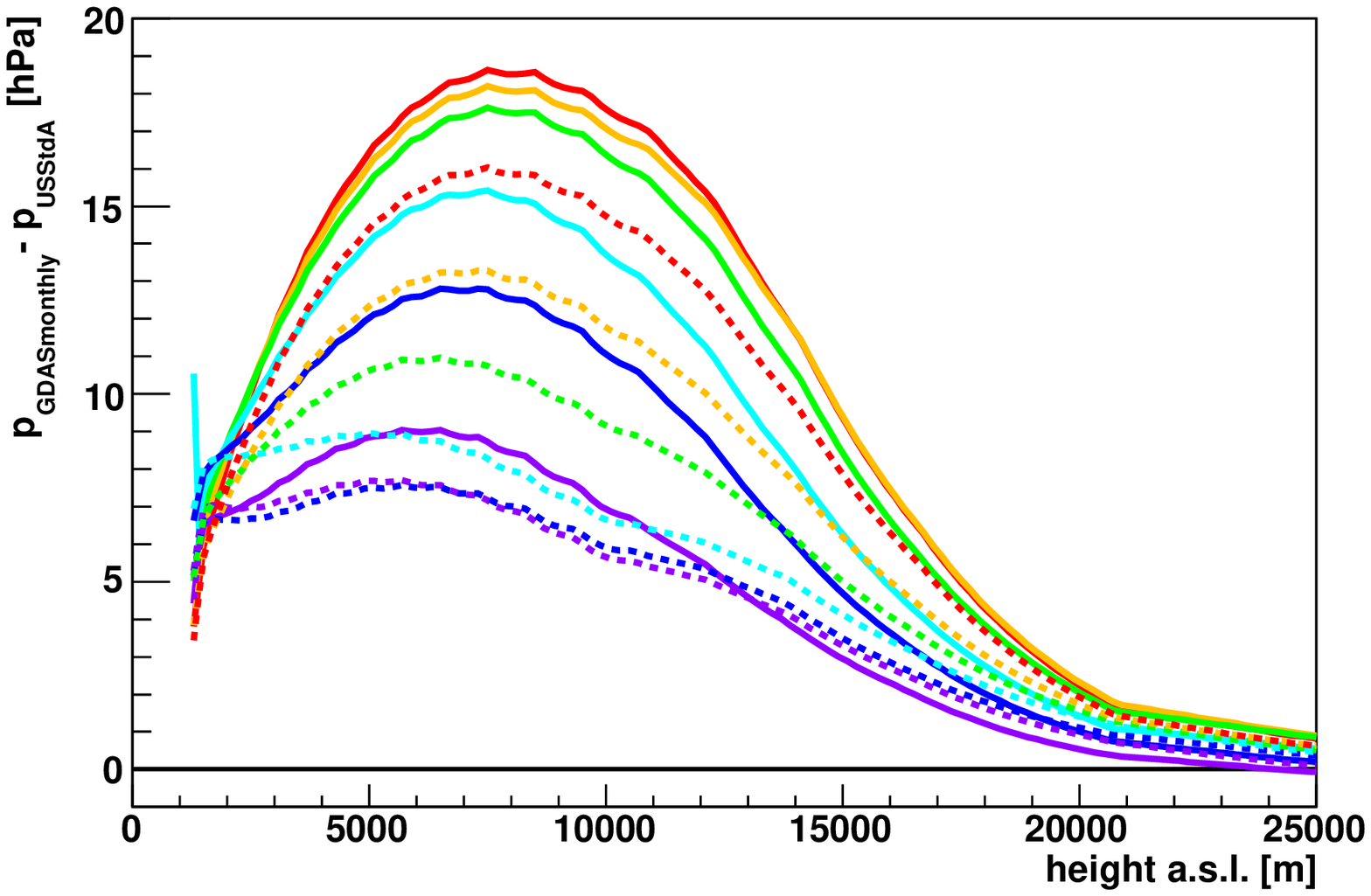}
  \end{minipage}
  \begin{minipage}[t]{.49\textwidth}
    \centering
    \includegraphics*[width=.9\linewidth,clip]{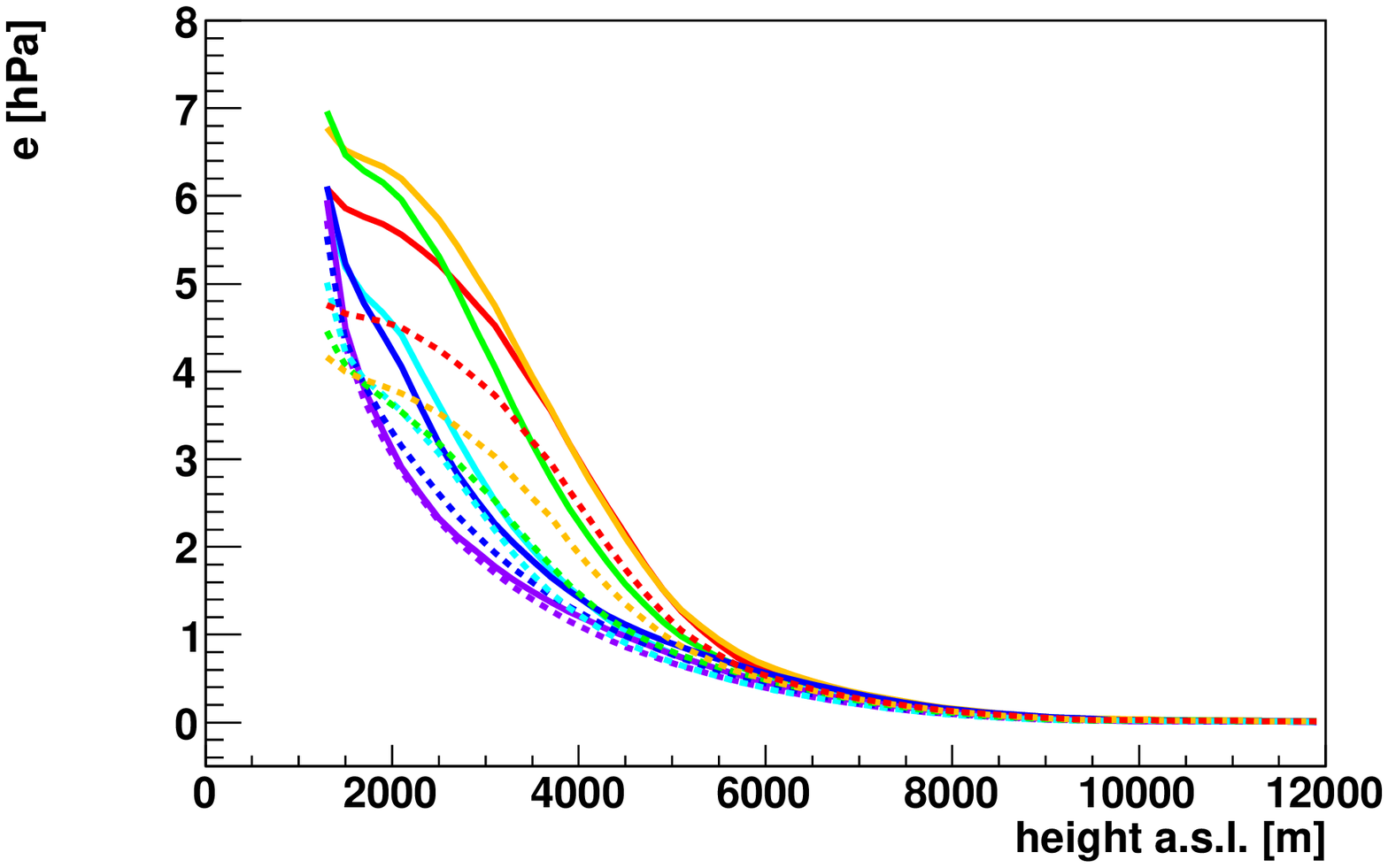}
  \end{minipage}
  \hfill
  \begin{minipage}[t]{.49\textwidth}
    \centering
    \includegraphics*[width=.9\linewidth,clip]{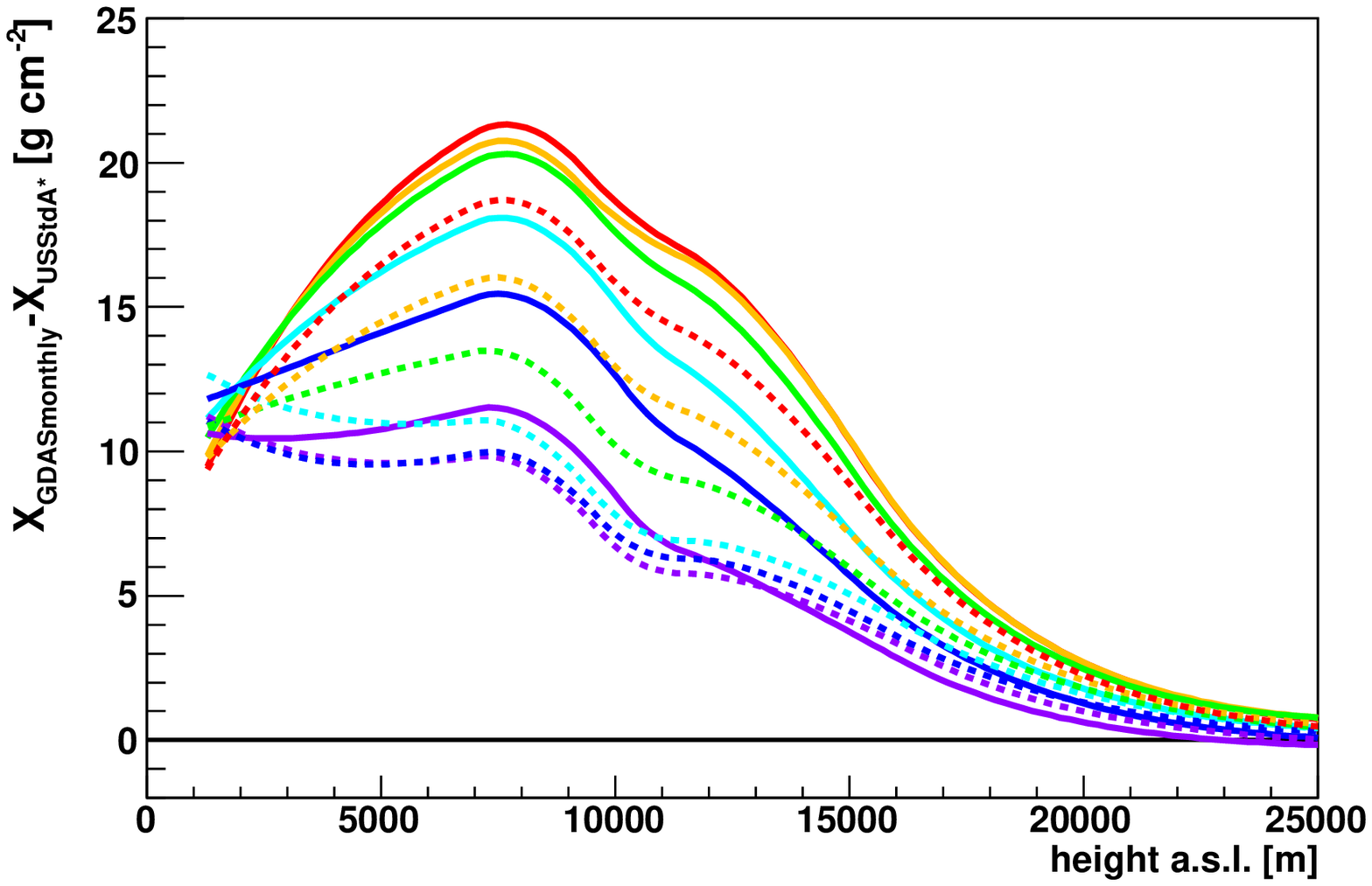}
  \end{minipage}
  \caption{\label{fig:means_final}
    Monthly averages after fit and self-consistency procedures for temperature
    $T$, pressure $p$, water vapor pressure $e$ and atmospheric depth $X$.
    Pressure and atmospheric depth are shown in difference to the US Standard
    Atmosphere.
  }
  \end{center}
\end{figure*}

For using the monthly GDAS models in air shower reconstructions as a fallback if
the 3-hourly GDAS data are not available, the models for the temperature,
pressure, humidity and air density have to be provided, along with the depth
parameterizations. However, after the fit procedure, the model of the
atmospheric depth is not necessarily consistent with the averages of the other
variables. Therefore, new air density profiles are computed from the derivatives
of the depth parameterizations. Using the averaged pressure profile, a new
temperature profile is calculated using the ideal gas law and new profiles of
the water vapor pressure and relative humidity can be compiled. The new profiles
are checked against the averaged profiles and the differences are found to be of
the order of the statistical uncertainties of the averaged models. In
Fig.~\ref{fig:means_final}, the final models for temperature, pressure, water
vapor pressure and the parameterizations of the atmospheric depth are shown.

\section{Reconstruction of Extensive Air Showers
\label{sec:reco}}

\begin{figure}[t]
  \begin{center}
    \begin{minipage}[t]{.49\textwidth}
      \centering
      \includegraphics*[width=.96\linewidth,height=.287\textheight]{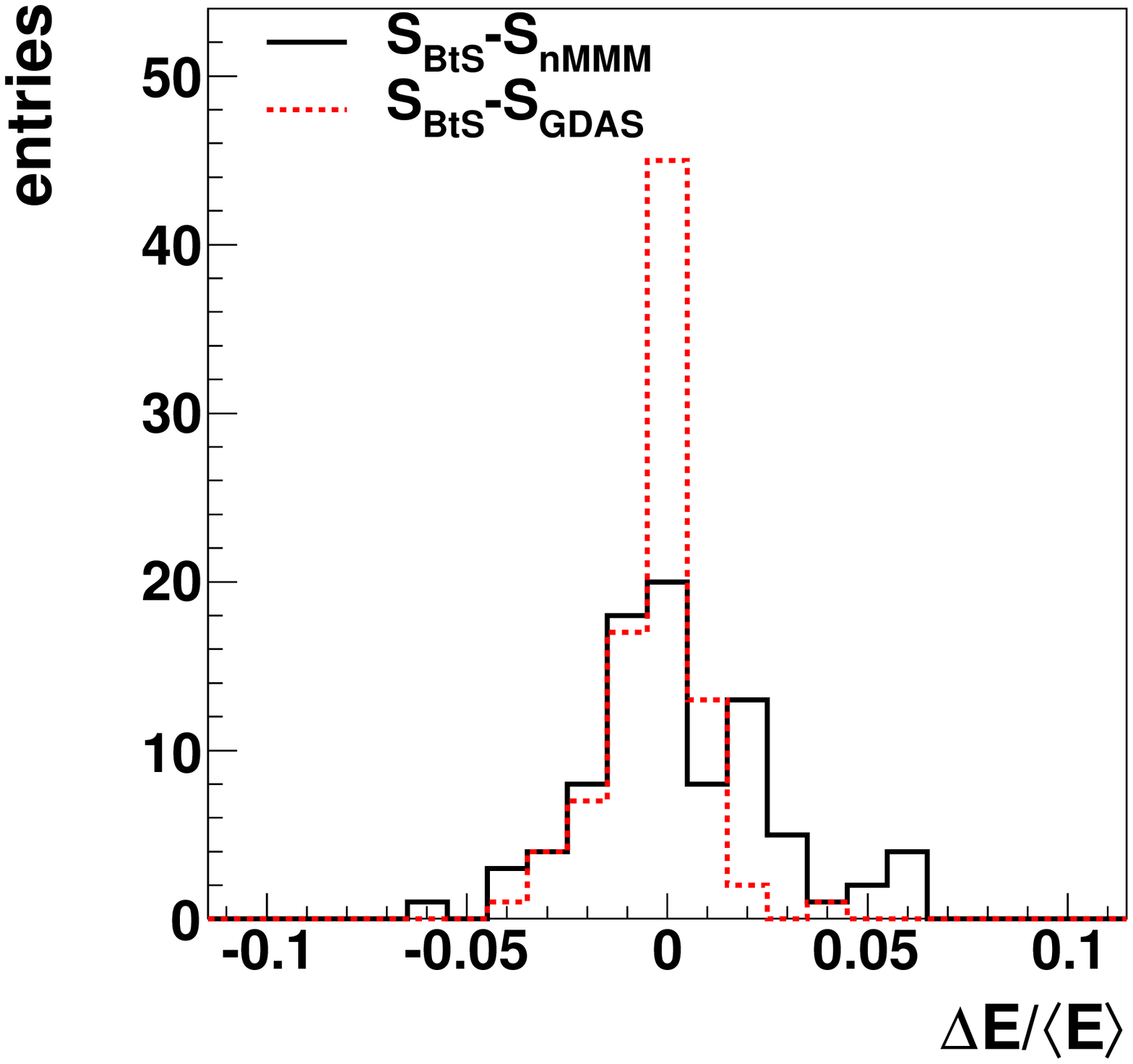}
    \end{minipage}
    \hfill
    \begin{minipage}[t]{.49\textwidth}
      \centering
      \includegraphics*[width=.96\linewidth,height=.287\textheight]{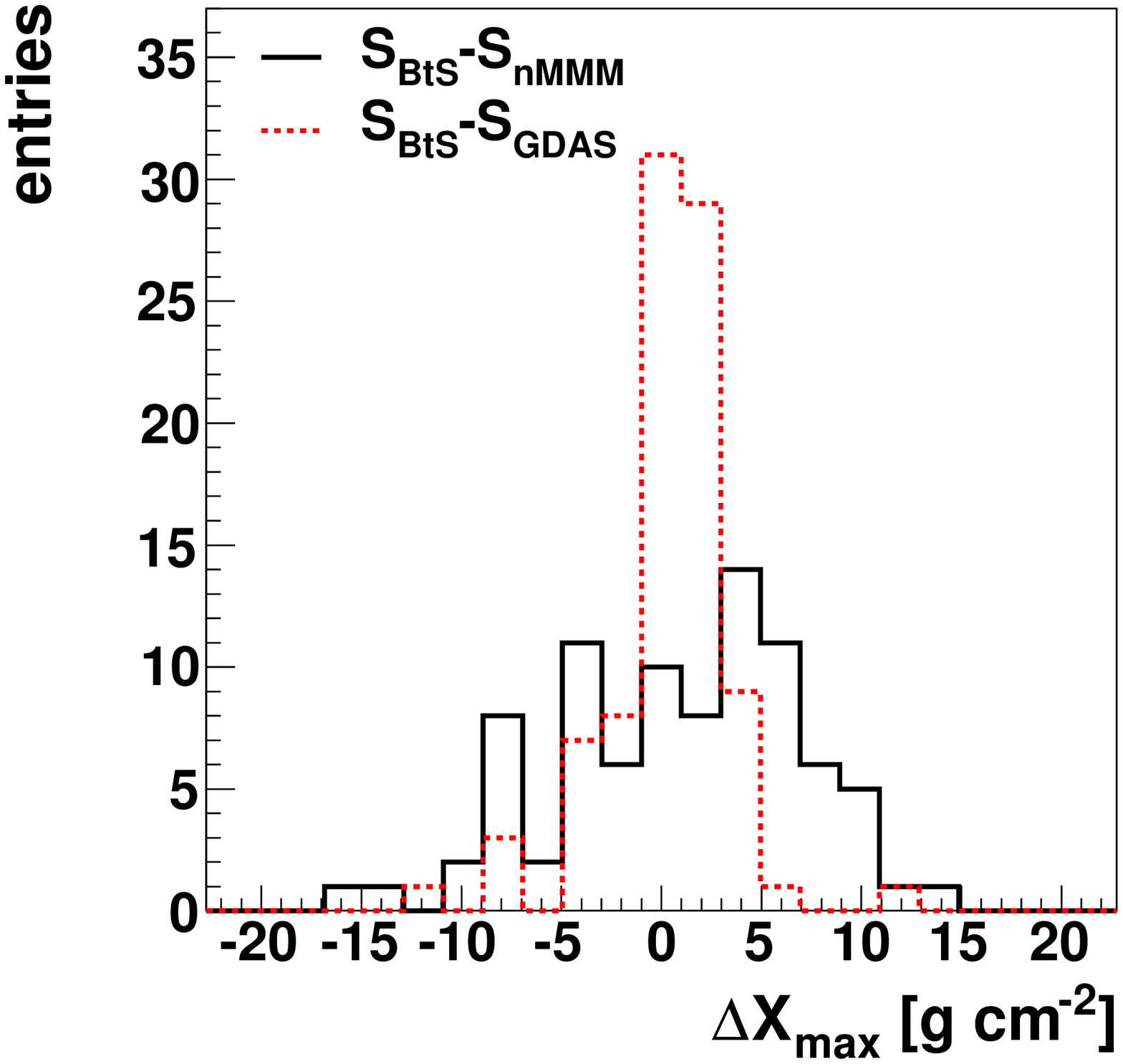}
    \end{minipage}
  \end{center}
  \caption{\label{fig:reco}
    Energy difference (left) and difference of position of shower maximum
    (right) between events recorded within the \bts program. The air showers are
    reconstructed with three different atmospheric descriptions, details see
    text.
  }
\end{figure}

As a final check of the applicability of the meteorological model data for the
measurements of the \pao, air shower data are reconstructed. In this analysis,
the two most important air shower parameters, primary energy $E$ and atmospheric
depth of shower maximum \xmax, are investigated. The same set of air shower
events is reconstructed in the same way with three different atmospheric
descriptions. To be able to use real-time atmospheric profiles, all air showers
of the \bts program are used. These 62 events are reconstructed one time with
the atmospheric profiles of the dedicated radio soundings, labelled \sbts. The
second reconstruction set is processed using the monthly models from local
measurements (Sec.~\ref{sec:nmmm}), \snmmm, and the third set is produced
applying the GDAS data with the 3-hour resolution, \sgdas. In
Fig.~\ref{fig:reco}, the differences for $E$ and \xmax are shown between the
different sets of reconstructions. For \sbts-\snmmm, the mean difference in the
reconstructed energy is $0.5\%$ and the width of the distribution is $2.3\%$, in
\xmax the difference is $0.3$\,\gcmsq with an RMS of $6.6$\,\gcmsq. For
\sbts-\sgdas, the mean difference in the reconstructed energy is $-0.2\%$ with
an RMS of $1.2\%$, and $0.5$\,\gcmsq with a width of $3.3$\,\gcmsq in \xmax.
While the mean differences are centered about zero, the spread of the
distribution of the \sbts-\snmmm comparison is not negligible for the
uncertainties of air shower reconstructions.

Applying the atmospheric data from GDAS presented here in an air shower
simulation study, the advantage of the data is demonstrated. The reconstruction
uncertainties related to the description of the molecular atmosphere decrease by
as much as 50\% using the GDAS data instead of the monthly
models~\cite{Abreu:2012gdas}. This result was obtained by simulating air showers
with arbitrary geometries using atmospheric conditions measured by radiosondes.
After reconstructing these data with the correct measured data, the GDAS data
and the monthly models, the uncertainties can be evaluated by comparing the
reconstructions using the models and the correct atmospheric conditions.

\section{Summary
\label{sec:summary}}

Atmospheric monitoring and the better understanding of fluorescence emission
processes in the atmosphere are very important tasks for ground-based cosmic ray
observatories, as they contribute a large part to the total systematic
uncertainties of reconstructed parameters.

At the \pao, atmospheric state variables at the ground are measured continuously
by several weather stations and an extensive weather balloon program was
operated to measure the height-dependent profiles of these parameters. To use
this information in the reconstruction of air showers, monthly models have been
constructed from monthly averages of these balloon data. To improve the
reconstruction quality of showers of special interest, a rapid monitoring
program was implemented to launch balloons after high-energetic events that are
identified by an online reconstruction.

Following the balloon program, GDAS data are now processed for using in the
analyses procedures of the \pao. With their superior temporal resolution and
good availability, they are suited perfectly for the reconstruction of air
showers and help to significantly reduce the systematic uncertainties associated
with the choice of the model to describe the molecular atmosphere. From these
data, monthly models were constructed to help to incorporate our best knowledge
of the atmosphere above the \pao also in the simulations of cosmic ray air
showers.

\section*{Acknowledgments}

We would like to thank the organizers of the workshop \emph{Interdisciplinary
Science @ the Auger Observatory: from Cosmic Rays to the Environment} in
Cambridge, UK, 2011, and in particular A.A.~Watson, for the inspiring meeting.
Part of these investigations are supported by the Bundesministerium f\"ur
Bildung und Forschung (BMBF) under contracts 05A08VK1 and 05A11VK1.
Furthermore, these studies would not have been possible without the entire
Pierre Auger Collaboration and the local staff of the Observatory.

\bibliography{gdas_epj}

\begin{thebibliography}{10}
\providecommand{\url}[1]{{#1}}
\providecommand{\urlprefix}{URL }
\expandafter\ifx\csname urlstyle\endcsname\relax
  \providecommand{\doi}[1]{DOI \discretionary{}{}{}#1}\else
  \providecommand{\doi}{DOI \discretionary{}{}{}\begingroup
  \urlstyle{rm}\Url}\fi

\bibitem{Abraham:2010}
J.~Abraham et~al., the Pierre Auger Collaboration, Astropart. Phys.
  \textbf{33}, 108 (2010)

\bibitem{Tonachini:2012lid}
V.~Rizi, A.~Tonachini, for the Pierre Auger Collaboration, in \emph{Focus Point
  on Interdisciplinary Science at the Pierre Auger Observatory (2012)},
  Editors: A.~Bueno, L.~Wiencke. EPJ Plus

\bibitem{BenZvi:2012xts}
P.~Abreu et~al., the Pierre Auger Collaboration, submitted to JINST  (August
  2012)

\bibitem{Wiencke:2012epj}
L.~Wiencke, for the Pierre Auger Collaboration, in \emph{Focus Point on
  Interdisciplinary Science at the Pierre Auger Observatory (2012)}, Editors:
  A.~Bueno, L.~Wiencke. EPJ Plus

\bibitem{Abreu:2012gdas}
P.~Abreu et~al., the Pierre Auger Collaboration, Astropart. Phys. \textbf{35},
  591 (2012)

\bibitem{Abraham:2009pm}
J.~Abraham et~al., the Pierre Auger Collaboration, Nucl. Instr. Meth.
  \textbf{A620}, 227 (2010)

\bibitem{Ave:2007b}
M.~Ave et~al., the AIRFLY Collaboration, Nucl. Instr. Meth. \textbf{A597}, 46
  (2008)

\bibitem{Abraham:2010b}
J.~Abraham et~al., the Pierre Auger Collaboration, Phys. Rev. Lett.
  \textbf{104}, 091101 (2010)

\bibitem{Arqueros:2008}
F.~Arqueros, J.~H{\"{o}}randel, B.~Keilhauer, Nucl. Instr. Meth. \textbf{A597},
  1 (2008)

\bibitem{Keilhauer:2008}
B.~Keilhauer, J.~Bl{\"u}mer, R.~Engel, H.~Klages, Nucl. Instr. Meth.
  \textbf{A597}, 99 (2008)

\bibitem{Arqueros:2011}
M.~Monasor et~al., Astropart. Phys. \textbf{34}, 467 (2011)

\bibitem{Keilhauer:2012}
B.~Keilhauer et~al., in \emph{International Symposium on Future Directions in
  UHECR Physics} (CERN, 2012)

\bibitem{Ave:2008}
M.~Ave et~al., the AIRFLY Collaboration, Nucl. Instr. Meth. \textbf{A597}, 50
  (2008)

\bibitem{Pereira:2010}
L.~Pereira et~al., Eur. Phys. J. \textbf{D56}, 325 (2010)

\bibitem{USStdA:1976}
{National Aeronautics and Space Administration (NASA)}, {U.S.\ Standard
  Atmosphere 1976}.
\newblock {NASA-TM-X-74335} (1976)

\bibitem{graw}
{GRAW Radiosondes GmbH \& Co. KG}, Tech. rep.
\newblock \urlprefix\url{http://www.graw.de}

\bibitem{Heck:1998}
D.~Heck et~al., {Forschungszentrum Karlsruhe Report} \textbf{6019} (1998)

\bibitem{Pesce:2011icrc}
R.~Pesce, for the Pierre Auger Collaboration, in \emph{Proc. 32nd ICRC}
  (Beijing, China, 2011)

\bibitem{Abraham:2007}
J.~Abraham et~al., the Pierre Auger Collaboration, Science \textbf{318}, 938
  (2007)

\bibitem{Abraham:2010yv}
J.~Abraham et~al., the Pierre Auger Collaboration, Phys. Rev. Lett.
  \textbf{104}, 091101 (2010)

\bibitem{Baus:2011icrc}
C.~Baus et~al., in \emph{Proc. 32nd ICRC} (Beijing, China, 2011)

\bibitem{GDASinformation}
{NOAA Air Resources Laboratory (ARL)}, {Global Data Assimilation System (GDAS1)
  Archive Information}.
\newblock Tech. rep. (2004).
\newblock \urlprefix\url{http://ready.arl.noaa.gov/gdas1.php}

\end{thebibliography}
\bibliographystyle{spphys}

\clearpage

\end{document}